\documentclass[aps,floatfix,twocolumn,showpacs]{revtex4}
\usepackage{amsmath}
\usepackage{threeparttable}
\usepackage{graphicx}
\usepackage{hyperref}
\def\etal{{\em et al. }}

\begin{document}
\title{Microscopic folding model analysis of the radiative $(n,\gamma)$ reactions near the $Z=28$ shell-closure and the weak s-process}
\affiliation{Department of Physics, University of Calcutta\\ 92, Acharya Prafulla Chandra Road, Kolkata-700009}
\author{Saumi Dutta}
\email{saumidutta89@gmail.com}
\author{G. Gangopadhyay}
\email{ggphy@caluniv.ac.in}
\author{Abhijit Bhattacharyya}
\email{abhattacharyyacu@gmail.com}
\begin{abstract}
The radiative thermal neutron capture cross sections over the range of thermal energies from 1 keV to 1 MeV are studied in statistical Hauser-Feshbach formalism. The optical model potential is constructed by folding the density dependent M3Y nucleon-nucleon interaction with radial matter densities of target nuclei obtained from relativistic-mean-field (RMF) theory. The standard nuclear reaction code TALYS1.8 is used for calculation of cross sections. The nuclei studied in the present work reside near the $Z=28$ proton shell closure  and are of astrophysical interests taking part in p-, s-, and r-process of nucleosynthesis. The Maxwellian-averaged cross-section (MACS) values  for energies  important for astrophysical applications are presented. 
\end{abstract}
\maketitle
 \section{Introduction}
After the termination of charged particle induced reactions, the stellar core contains mainly iron and a few trans-iron elements. These elements can act as seeds and take part in further nucleosynthesis by capturing neutrons. Depending upon the conditions of stellar temperature and neutron density, two different nucleosynthesis mechanisms can be possible. The slow neutron capture process or s-process, in which, the $\beta$-decay rates exceed the neutron capture rates, becomes active at relatively low neutron density and temperature, in which, the elements are formed near the valley of $\beta$-stability. On the other hand, the rapid neutron capture process or r-process operates at very high neutron density, in which, the neutron capture rates remain very high compared to $\beta$-decay rates. A minor contribution comes from a different process, known as, 
p-process, which drives the material through the proton drip line of the nucleosynthesis chart, either by photo-disintegration, basically, the $(\gamma,n)$ reactions, or by capturing protons. The nuclei thus produced, are termed as the p-only nuclei or simply the p-nuclei.

 While the majority of the theory of s-process has developed, there are still some questions those have to be addressed. The determination of the exact nucleosynthesis path requires a network calculation, the key inputs of which are the thermal neutron capture rates. The  s-process is subdivided into weak, main, and strong components. The weak component is dominated in between Fe to Sr-Zr-Mo region. 

The weak s-process, which is responsible for most of the s-abundances in the mass region $A=60-100$, occurs in massive stars (M$\geq$ 8$M_{\odot}$, $M_{\odot}$ is the solar mass) \cite{weak_sp}. The neutron exposure in massive stars is not too high and hence, the  local approximation, i.e., $\sigma$N$_{s}$=constant, does not hold good. Hence, the abundance pattern suffers from strong propagation effects for cross-section uncertainties. The uncertainty in the nuclear cross section of a single isotope does not affect the abundance of that particular isotope only but influences the abundances of subsequent isotopes or the entire distribution. Hence, accurate cross sections are necessary to eliminate the strong propagation effects in the abundance pattern.  
  
The isotopes in the neutron-rich side, those rely on outside the s-process path, are ascribed to the r-process and are called r-only nuclei. On the other hand, those, which are shielded against the r-process $\beta$-decay flow by the stable isobars, are produced only in the s-process and are termed as the s-only isotopes. In an explosive r-process scenario, when the freeze-out is achieved, the whole material $\beta$-decays back towards the valley of stability and hence, they mix with the s-abundances.
Many isotopes have origins from a complex mixture of the p-, s-, and r-processes \cite{psr}. The origin of these elements and their formation by different processes in accurate proportion are being studied in recent years. Hence, total $(,\gamma)$ cross sections are crucial inputs in this respect. 

 Some nuclei in this region act as bottlenecks due to their small cross sections. There are a few branching where the rates of  $\beta$-decay and neutron capture become comparable. Various information regarding stellar conditions during the nucleosynthesis  can be drawn from the analyses of branching. For example, one can estimate the neutron density, temperature, electron density, etc., from an accurate analysis of branching. 
At very high neutron density, many branching, which in general are not considered in classical s-process network calculation, can be active. Hence, one has to construct a much larger network in such conditions.  In the weak region in massive stars, this process is termed as weak sr-process \cite{sr-process}. For example, a significant amount of $^{60}$Fe can be produced during the high neutron flux in shell carbon burning phase of s-process \cite{fe60_1}. This is the result of branching at $^{59}$Fe with the half-life of  $44.495$ days. At low neutron densities, neutron capture rate on $^{59}$Fe remains lower than the corresponding  $\beta$-decay rate and the branching does not occur.  Thus, one must have complete and proper knowledge of neutron capture cross sections of unstable isotopes in order to study such cases. In such scenarios, theoretical calculations remain the only way to predict the values.  

Furthermore,  a few isotopes lie in the vicinity of  the bridging region of  weak and main components of the s-process. Hence, besides  their production in the weak s-component in massive stars, with masses greater than 8$M_{\odot}$, a  small fraction of these isotopes is produced in AGB stars. For example, the main s-process accounts for $\sim$5\% of solar copper and $\sim$10\% of solar gallium, germanium, and arsenic \cite{book}. Hence, their cross sections are required for the discrimination between the weak and the main s-component contributions.

 Precise capture rates have also consequences for r-process study.
Once accurate s-abundances are obtained, one can easily find abundances of the solar r-process residuals 
 by simply subtracting from total solar abundance.  
Some of the isotopes in this region of interest are important for galactic chemical evolution \cite{galactic}. 
Some studies have revealed overabundance problem
 of certain elements in this region of interest. An explanation of this is hindered due to the limitations in the accurate cross sections.
Apart from the astrophysical point of view, the trans-iron elements are also important as structural materials for nuclear reactor applications. Hence, from this perspective also, the neutron cross sections are significant.  

We have calculated the radiative thermal neutron capture cross sections for  nuclei near the $Z=28$ proton shell closure from a theoretical viewpoint with the statistical reaction code TALYS1.8 \cite{talys}. The aim of our work is to set up a definite and consistent theoretical model that can efficiently predict the cross sections  for nuclei over a large mass range as well as energy range for astrophysical applications. This will supplement those cases for which measurement is not possible or not yet done. The paper is organized as follows. In the next section, we  briefly describe the theoretical formalism of our work. In section III, we have discussed results the $(n,\gamma)$ cross sections after comparing them with the available experimental data. Then Maxwellian-averaged cross section (MACS) values  are presented for the nuclei, at an energy of 30 keV.  We have also given the MACS values over a range of energies  useful for stellar model calculations for a few nuclei which do not have any experimental data.
 Lastly, the summary  is presented.

\section{Theoretical formalism}

In the case of neutron-induced reactions, a neutron as a projectile, upon incidenting on several target nucleus results in a binary reaction $A(a,b)B$. The target and the neutron together forms a compound nucleus  with a total energy $E_{tot}=E_{cm}+S_{n}+E_{x}$, and a range of values of spins (J) and parities ($\pi$). Here, $E_{cm}$ is the incident energy in the center of mass frame, $S_{n}$ is the  neutron separation energy of compound nucleus, and $E_{x}$ is the excitation energy of the target which is zero when the target is in the ground state.  
In the present work, we are studying the radiative neutron capture, i.e., $(n,\gamma)$ reactions where the compound nucleus, after its formation, decays back 
to its ground state by emitting subsequent $\gamma$-rays.

In the very basic sense, the neutron capture cross section $\sigma_{n,\gamma}$, which is a measure of probability of neutron capture, is an effective area that the target presents before the neutron for its absorption and is defined as the ratio of the number of reactions occurred per unit time per target nucleus  to the total incident flux of incoming neutrons. The cross section is dependent on energy of the system and is a sum of compound nuclear term which is described by Hauser-Feshbach formula, individual resonances that are determined from Breight-Wigner formula, the direct capture components those are proportional to $1/v$, $v$, $v^{3}$, etc. for s, p, d-waves, respectively and so on, and to a certain extent, the interference between direct capture and single resonances. Resonances are observed at low excitation energies when the separation between the levels is large so that the individual peaks appear in the cross section. However, in general, in the statistical model calculation of reaction cross sections in Hauser-Feshbach formalism assumes a large number of resonances at compound formation energies so that the individual resonances can be averaged over the closely spaced overlapping levels. 
This cross section is defined as,
\begin{equation}
\sigma_{HF}=\sigma_{form}\frac{\Gamma_{\gamma}}{\Gamma_{tot}}
\end{equation}
Here, $\sigma_{form}$ is the formation or the absorption cross section of the compound nucleus.
 $\Gamma_{\gamma}$  is the partial decay width to $\gamma$ channel and $\Gamma_{tot}$ is the total decay width of all possible exit channels.
It is true that the intermediate mass nuclei near the closed shells do not have high
 level density and hence, most of the statistical model calculation fails near the 
closed neutron or proton shells. It is our aim to test the validity and reliability 
of our constructed model in the prediction of reaction cross sections 
 near the magic numbers.
The definition of compound nuclear contribution to the total cross section according to Hauser-Feshbach formula in a compact form can be given as \cite{iliadis},
\begin{equation}
\sigma_{ab}=\frac{\pi}{k_{\alpha}^{2}} \sum_{J \pi}\frac{(2J+1)}{(2I_{1}+1)(2I_{2}+1)}
\frac{T_{a}T_{b}}{\sum_{c}T_{c}}W_{ab}
\end{equation}
for each combination of a and b, where the set $a=\{\alpha,l,j\}$ and the set $b=\{\alpha^{\prime},l^{\prime},j^{\prime}\}$. The unprimed and primed quantities are for incident and outgoing channels, respectively. Here, $l$, $s$, and $j$ denote the orbital angular momentum, spin, and total angular momentum, respectively where as $\alpha$ and $\alpha^\prime$ are the channel designators for  projectile$+$target system and residual nucleus$+$ejectile system. The average transmission coefficients for incident and outgoing channels are denoted by $T_{a}$ and $T_{b}$  while $T_{c}$ denotes average transmission coefficients for compound system. $I_{1}$, $I_{2}$, are the spins of target and projectile, $J$ is the total angular momentum of the compound nucleus.
At low incident energies ($\textless$ 1 MeV) and for medium mass targets, 
especially, when it is lower than the threshold excitation energy of the first inelastic level,
  elastic scattering and radiation capture are the dominating processes over 
inelastic scattering or other reaction channels those  gradually open up at higher energies. $W_{ab}$ is the width fluctuation correction factor. These are the crucial renormalization factors  to conserve the average cross section. For example,
 it may be possible that the emission of ejectile occurs at very early stage of compound nuclear formation before the equilibration or redistribution of energies over all states in the compound system via a sufficient number of collisions takes place. This results in strengthening the elastic scattering channel over the others and a renormalization of each transmission coefficient in the outgoing channel have to be performed accordingly for the appropriate quantitative description of cross sections. This effect is  especially severe near the threshold energies of new channel openings where the channel strengths differ significantly and for low projectile energies when only a few channels exist in outgoing part.

The entrance channel has neutrons and hence, neutron transmission coefficient directly enters into the calculation. 
These are obtained from a complex optical model potential that can describe the reaction via its imaginary part.
 The optical model potential describes complicated many-body nucleon-nucleus interaction by an average one body potential. The wave functions for both elastic scattering and reactions can then be obtained by solving the Schr\"odinger equation with this complex potential and then, from the phase shifts one can easily determine the transmission coefficients. Thus, the optical model provides the basis for the theoretical calculation of  cross sections those can be utilized in various practical applications. Although the earlier approach of phenomenological potential with a large number of parameters those are adjusted to fit the experimental measurements is successful, it is only limited to those regions where sufficient amount of experimental information are available to constrain its parameters. In this regard, recently more accurate microscopic models are being developed by folding the nuclear matter densities with the inherently complicated nucleon-nucleus interaction. The basic advantage of such microscopic models is that they can be reliably applied to regions far from the nuclear stability valley. 
  We have constructed a microscopic neutron optical potential from density-dependent M3Y (DDM3Y) nucleon-nucleus 
interaction \cite{ddm3y}, based on G-matrix oscillator basis. This interaction is then folded with target radial matter densities, obtained from relativistic mean field calculation. The folding is done in coordinate space with spherical symmetry. The folded potential in MeV  is given as,
\begin{equation}
V_{fold}({\bf r}, E)=\int v|({\bf r}-{\bf r\prime}, \rho, E)| \rho({\bf r\prime}) d{\bf r\prime}
\end{equation}
The interaction $v({\bf r}, \rho, E)$ contains 
two direct terms of different ranges
 according to distinct nature of nuclear force and an energy dependent zero-range pseudo-potential
 component  representing exchange term. The interaction in MeV is as follows,
\begin{equation}
\begin{aligned}
v({\bf r}, \rho, E)=2.07\biggr[7999\frac {e^{-4r}}{4r}-2134\frac{ e^{-2.5r}}{2.5r}
&&\\-276\biggr(1-0.005\frac{E}{A}\biggr) \delta(r)\biggr]\biggr(1-1.624 \rho^\frac{2}{3}\biggr) 
\end{aligned}
\end{equation}

Here, $E$ is the projectile energy in the center of mass frame.
The folded DDM3Y potential serves as the real part. The optical model potential has been formulated by taking its imaginary part identical to the real part and finally a renormalization has been done by multiplying both the real and imaginary components by numerical  factors.
\begin{equation}
\label{norm_eqn}
V_{omp}=A_{r}V_{fold}+A_{im}V_{fold}
\end{equation}
Earlier, this optical potential was found to describe proton capture reactions over a wide ranges of mass of targets
 \cite{55-60,60-80,80-90,110-125,40-55}.

The relativistic mean field model, used to obtain baryonic matter density, is based on FSU Gold parameterization \cite{fsugold,n82}. The mesonic part of the lagrangian  contains fields for isoscalar-scalar $\sigma$ meson, isoscalar-vector  $\omega$ meson, and isovector-vector $\rho$ meson. Apart from the usual couplings between the nucleon field and meson fields, this RMF model, in addition, contains  nonlinear meson self-interaction terms
% that can reliably soften the EOS of  the symmetric nuclear matter.
 Pairing is incorporated in the continuum BCS approximation using a 
delta pairing potential $V({\mathbf r1}, {\mathbf r2})=-V_{0}\delta(\bf{r1}-\bf{r2})$, where the pairing strength $V_{0}$ has been chosen to be 300 MeV for both protons and neutrons. No-sea approximation has been used, i.e., the contribution of baryons from the vacuum has not been considered.
%The field equations have been solved in coordinate space in iterative technique with spherical approximation using  Runge-Kutta method and  Simpson's rule of fourth order for Klein-Gordon equations for meson fields and Dirac equations for nucleon fields, respectively.

Further, we have convoluted the point proton densities with the standard Gaussian form factor F({\bf r}) to obtain the charge distribution and then the root-mean-square (rms) charge radius values in order to check the validity of the RMF model used. 
\begin{equation}
\rho_{ch}({\bf r})=e\int \rho_{p} ({\bf r\prime})F({\bf r}-{\bf r\prime}) d {\bf r\prime}
\end{equation}
\begin{equation}
F(r)=(a \sqrt\pi)^{-3}exp(-r^{2}/a^{2})
\end{equation}
with $a=\sqrt{2/3}a_p$, where $a_p=0.80$ fm is the root-mean-square (rms) charge radius of  the proton. 
\begin{equation}
R_{rms}=\sqrt \frac{ \int r\prime^{2}\rho_{ch}({\bf r\prime}) d{\bf r\prime}}{\int \rho_{ch}({\bf r\prime}) d{\bf r\prime}}
\end{equation}

Photon transmission coefficient is one of the crucial inputs as $\gamma$-transmission is the dominating channel
for nuclear deexcitation at energies below few MeVs, especially for neutron-induced reactions. The $\gamma$-ray
 transmission coefficient for multipolarity $l$ and $\gamma$-ray energy $E_{\gamma}$ for type X 
(stands for electric $(E)$ or magnetic $(M)$ is given by,
\begin{equation}
T_{Xl}(E_{\gamma}) = 2\pi f_{Xl}(E_\gamma)E_\gamma^{2l+1} 
\end{equation}

Obviously, the leading contribution comes from the electric dipole $(l=1)$  transition,
 for which $T_{E1}$ is  essentially proportional to $E_{\gamma}^{3}$. Since the $\gamma$ transmission coefficient calculation involves all the possible states to which a photon
can be emitted from the initial compound nucleus state, the number of radiative open channels is almost
infinite, but each has a very small transmission coefficient. 
Here, $f_{Xl}(E_\gamma)$ is the energy-dependent $\gamma$-ray strength function. Theoretical predictions are necessary due to the incompleteness of the  experimental database. There are several methods available for the calculation of $\gamma$-ray strength function. The realistic phenomenological closed form models such as standard Lorentzian model, the hybrid model, generalized Fermi liquid model, etc are 
gradually replaced by microscopic models those are correlated with nuclear structure properties, due to their superiority in predictive power. Moreover, the phenomenological models suffer from certain severe shortcomings. The predicted values are ambiguous or inappropriate for exotic nuclei and at energies around the neutron separation energy. In our present study, we have taken the values of E1 $\gamma$-ray strength function from the microscopic Hartree-Fock-Bogolyubov (HFB) + QRPA calculation of Goriely \etal \cite{strength_4} from drip line to drip line. It took into account the pairing effects and collective excitations.  The QRPA strength was folded with a
 Lorentzian distribution to generate the experimentally observed GDR widths. The widths were 
then modified in the framework of second RPA.

One of the important inputs in statistical  calculations are the nuclear level densities which are used whenever the information about the discrete level is not available. Nuclear level density (NLD) is  the number of nuclear levels per energy interval around an excitation energy, for a certain spin and
parity. 
Experimental information on NLD is limited to only low excitation energies and to those nuclei which are terrestrially accessible for measurement. However, for specific  applications, for example in the astrophysical studies involving nuclei along neutron or proton drip line, it is required to extrapolate the data in large extent far beyond the experimentally known region. Therefore, for large-scale applications, data have to be taken from  reliable theoretical models as it has been observed that the largest uncertainty in statistical model calculations stems from inappropriate description or prediction of NLDs. Hence, it is of prime interest to choose a physically sound theoretical model. Nowadays, there have been major improvements in deriving microscopic  models  over the earlier empirically adjusted  phenomenological models. We have taken the data from the recently developed microscopic model of Goriely \etal \cite{ldmodel_4} in the combinatorial method including collective rotational and vibrational phonon enhancements to predict spin-, parity-, and energy-dependent  NLDs. Goriely {\it et al.} used the Boson partition function \cite{boson_part_func}, the Hartree-Fock Bogolyubov ground state properties \cite{hfb_gr}, and BSk14 interaction \cite{bsk14}.

For neutron-induced reactions, the energy range and position of the peak of distribution  is governed by the centrifugal quantum number $l$ and hence, in general, is determined from the contribution of various partial waves. According to a simple approximation, the peak ($E_{0}$) and width ($\Delta$) in MeV are obtained as,
\begin{eqnarray}
E_{0}=0.172T_{9}\left(l+\frac{1}{2}\right)\\
\Delta=0.194T_{9}\left(l+\frac{1}{2}\right)^\frac{1}{2}
\end{eqnarray}
Hence, for pure s-wave neutron interaction, the peak coincides with that of Maxwellian-Boltzmann distribution function.

Neutrons in the interstellar medium are thermalized due to a  large number of collisions and there is obviously a thermal distribution of neutron velocity. It is, therefore, necessary to have the knowledge of average values of cross sections by folding them with distribution function. 
In the  high-temperature and high-density stellar plasma, quantum effects are negligible and hence, classical Maxwellian-Boltzmann (MB) distribution for neutron velocity is a good approximation. The Maxwellian-averaged cross sections (MACS) are obtained by folding the total $(n,\gamma)$ cross section with the MB distribution function. These MACS values are generally used in the quantitative calculation of abundances during various phases of evolution of the astrophysical medium. For s-process studies in the classical or canonical scenario, a single MACS at 30 keV are demanded. However, more general network calculations coupled with stellar codes that take into account the temporal  evolutionary phases of dense stellar matter require MACS values over a range of neutron energy. Experiments are not possible  at all energies and hence, 
theoretical extrapolations are evidently needed.  In our earlier studies, this theory was found to be successful in the study of the neutron capture reactions  for several nuclei those take part in heavy element nucleosynthesis near the $N=82$ as well as the $N=50$ shell closures \cite{n82,n50}. Some more details on theoretical description can also be available there.

\section{Results}
\subsection{Relativistic-Mean-Field Results}
\begin{figure*}
%\center
\includegraphics[scale=0.0650]{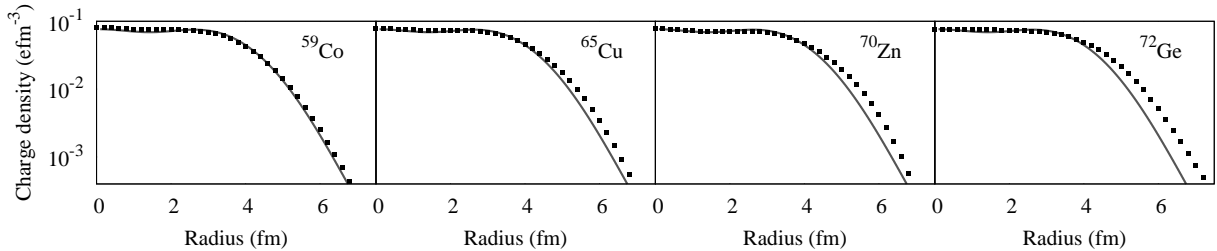}
\caption{Charge density profiles of $^{59}$Co, $^{65}$Cu, $^{70}$Zn, and $^{72}$Ge, from the relativistic mean field theory, is compared with the Fourier-Bessel parameter fit to the elastic electron  scattering data, taken from DeVries \etal \cite{devries}. The solid lines represent theoretical results and the discrete points represent the Fourier-Bessel parameter fit to the scattering data.
\label{dens_profile}}
\end{figure*}

\begin{table}[hbt]
\center
\caption{Rms charge radius values, extracted from relativistic-mean-field theory,
 are compared with the experimental data for the stable nuclei, studied in the present work. Experimental values are from Ref. \cite{angeli}. 
\label{chrad}}
\begin{tabular}{crc|crr}\hline
Nucleus & \multicolumn{2}{c}{Charge radius (fm)}&Nucleus&\multicolumn{2}{c}{Charge radius (fm)}  \\\hline
   & Present & Experiment&&Present&Experiment\\
\hline
$^{56}$Fe&3.6936&3.7377&$^{57}$Fe&3.7073&3.7532\\
$^{58}$Fe&3.7211&3.7745&$^{59}$Co&3.7505&3.7875\\
$^{58}$Ni&3.7497&3.7757&$^{60}$Ni&3.7777&3.8118\\
$^{61}$Ni&3.7912&3.8225&$^{62}$Ni&3.8113&3.8399\\
$^{64}$Ni&3.8257&3.8572&$^{63}$Cu&3.8467&3.8823\\
$^{65}$Cu&3.8647&3.9022&$^{64}$Zn&3.8775&3.9283\\
$^{66}$Zn&3.8917&3.9491&$^{67}$Zn&3.8986&3.9530\\
$^{68}$Zn&3.9056&3.9658&$^{70}$Zn&3.9366&3.9845\\
$^{69}$Ga&3.9486&3.9973&$^{71}$Ga&3.9688&4.0118\\
\hline
\end{tabular}
\end{table}
First we have presented the results of our RMF calculations.
 In Table \ref{chrad}, we have compared rms charge radius values of nuclei in the present study with the measured values. The experimental data are taken from I. Angeli \cite{angeli}.  Fig.\ref{dens_profile} shows the radial charge  density profiles of some selected nuclei in the region of interest. 
It can be seen that the RMF theory  reproduces the measurements very well.

\begin{figure*}
\center
\includegraphics[height=2.5in,width=5.7in]{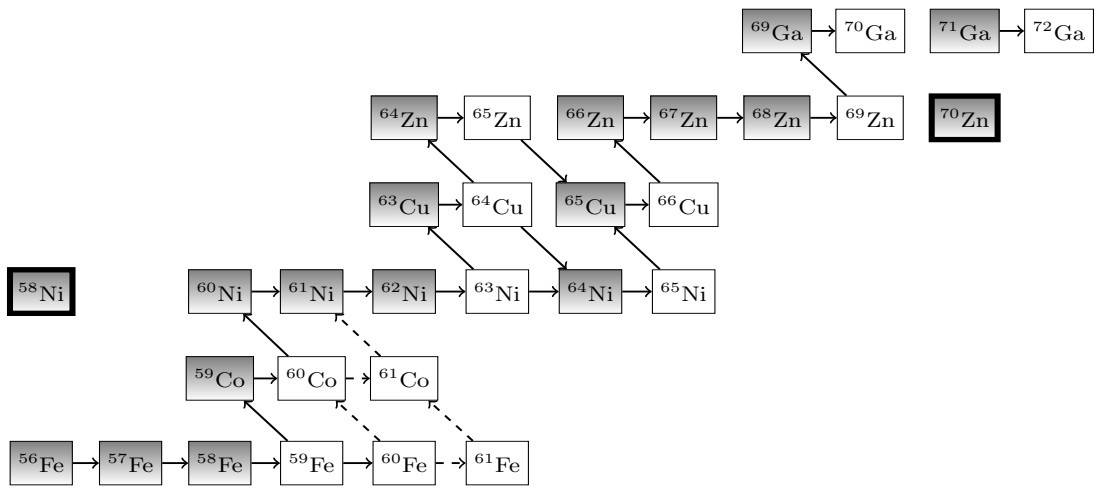}
\caption{Nucleosynthesis path from Fe to Ga. Shaded rectangles represent the stable isotopes. The p-only  and r-only nuclei are designated by rectangles with thick borders. 
\label{path}}
\end{figure*}

\subsection{The neutron capture  cross sections}
\begin{figure}
\center
\includegraphics[scale=0.60]{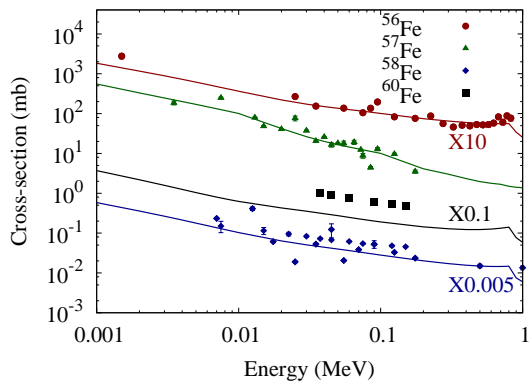}
\caption{ Comparison of $(n,\gamma)$ cross sections of the present calculation
with experimental measurements for $^{56,57,58,60}$Fe. The solid lines indicate the theoretical results. For the convenience of viewing, we have multiplied the cross sections of $^{56,58,60}$Fe by  factors of 10, 0.005, and 0.1, respectively. 
\label{fengxs}}
\end{figure}

\begin{figure}
\center
\includegraphics[scale=0.60]{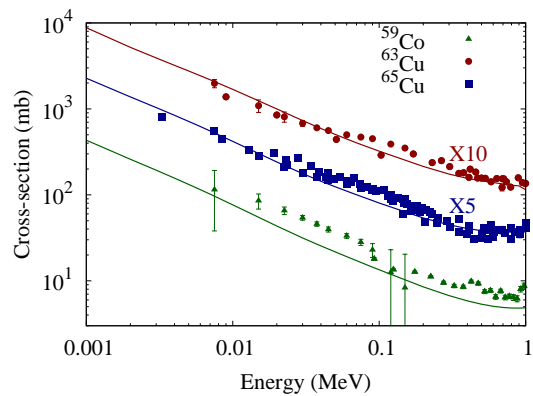}
\caption{Comparison of $(n,\gamma)$ cross sections of the present calculation
with experimental measurements for $^{59}$Co and $^{63,65}$Cu. The solid lines indicate the theoretical results. For the convenience of viewing, we have multiplied the cross sections of $^{63}$Cu and $^{65}$Cu by  factors of 10 and 5, respectively. 
\label{cucongxs}}
\end{figure}

\begin{figure}
\center
\includegraphics[scale=0.60]{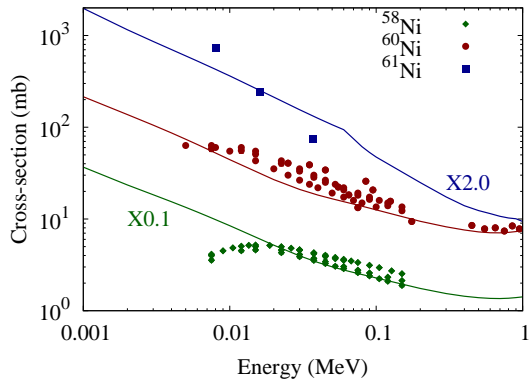}
\caption{ Comparison of $(n,\gamma)$ cross-sections of the present calculation
with experimental measurements for $^{58,60,61}$Ni. The solid lines indicate the theoretical results. 
For the convenience of viewing, we have multiplied the cross-section of $^{58}$Ni and $^{61}$Ni by factors of 0.1 and 2.0, respectively. 
\label{ningxs1}}
\end{figure}

\begin{figure}
\center
\includegraphics[scale=0.60]{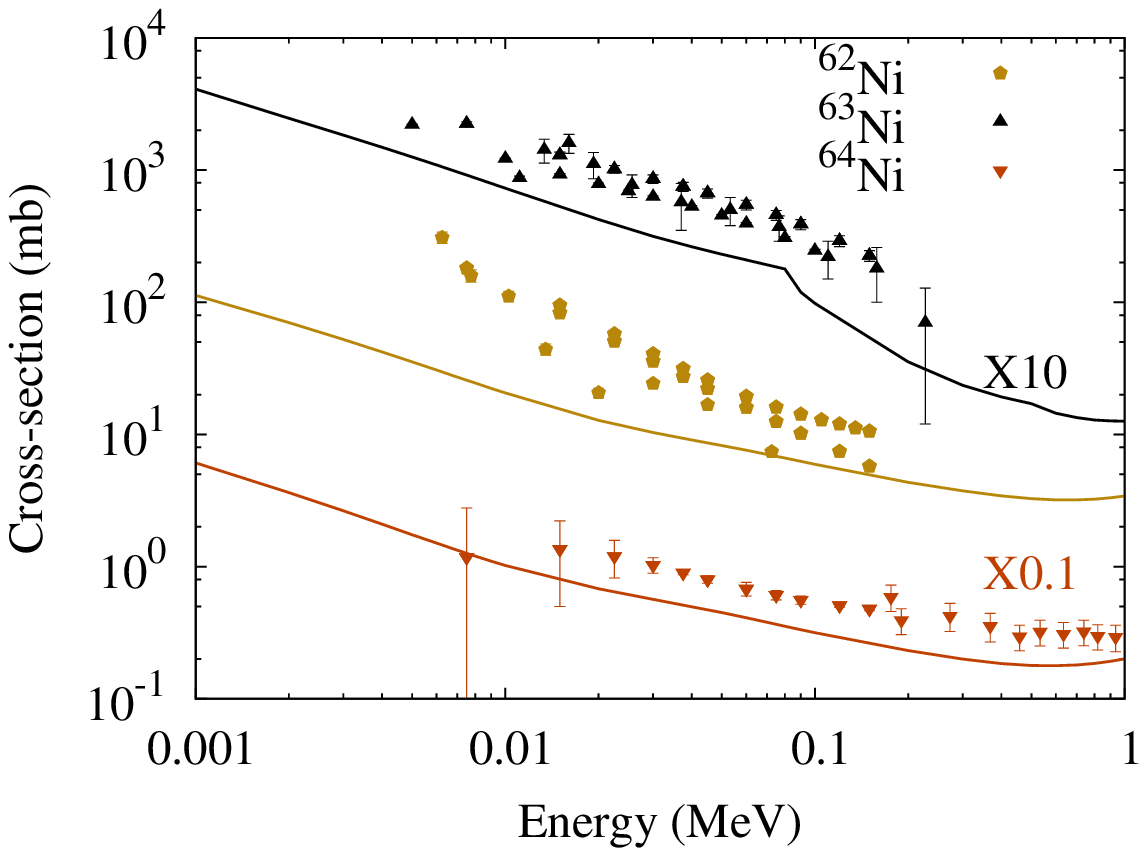}
\caption{Comparison of $(n,\gamma)$ cross sections of the present calculation
with experimental measurements for $^{62,63,64}$Ni. The solid lines indicate the theoretical results. 
For the convenience of viewing, we have multiplied the cross sections of $^{63}$Ni and $^{64}$Ni by a factors of 10 and 0.1, respectively. 
\label{ningxs2}}
\end{figure}

\begin{figure}
\center
\includegraphics[scale=0.60]{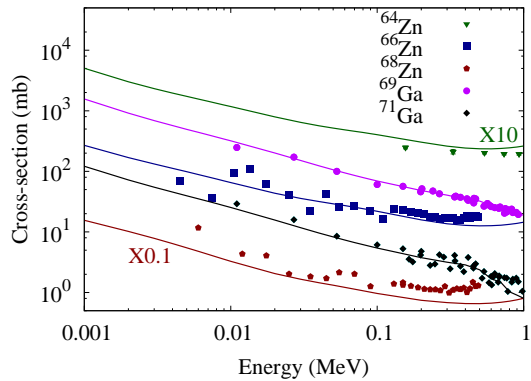}
\caption{Comparison of $(n,\gamma)$ cross sections of the present calculation
with experimental measurements for $^{64,66,68}$Zn and $^{69,71}$Ga. The solid lines indicate the theoretical results.
 For the convenience of viewing, we have multiplied the cross sections of $^{64}$Zn and $^{68}$Zn by factors of 10 and 0.1, respectively. 
\label{zngangxs}}
\end{figure}

Theoretical neutron capture cross sections as a function of neutron energies are compared with existing experimental data in Figs. \ref{fengxs}-\ref{zngangxs}. The old measurements, in general, did not have the facilities  of modern  improved techniques. For example, experiments on  $^{66,68}$Zn and $^{69}$Ga were performed more than 30 years ago. Thus, they suffer from large uncertainties. Recently, Heil \etal \cite{heil2008} carried out activation measurements on $^{58}$Fe, $^{59}$Co, $^{64}$Ni, and $^{63,65}$Cu with repeated irradiations at thermal energy of 25 keV. From the neutron spectrum, they have finally derived the MACS values after normalizing the measured data with existing differential cross-section values of the data libraries. Furthermore, most of the nuclei in this region have cross sections less than 100 mb. Hence, the smaller the cross sections, the greater is the probability of errors in the measurements and careful techniques have to be employed.  Thus, the impact of propagation effect is also expected to be more severe over the abundance distribution. 

The nucleosynthesis path in the region from Fe to Ga is shown in Fig. \ref{path}. The stable and extremely long-lived radionuclides are shown by shaded rectangles. The p-only isotope $^{58}$Ni and r-only isotope $^{70}$Zn are denoted by rectangles with thick borders. 

  Fig. \ref{fengxs}  shows the neutron capture cross sections of $^{56-60}$Fe.
 The iron nuclei act as seed elements in the s-process nucleosynthesis.
 The experimental data are from Refs. \cite{fe56_1,fe56_2fe57_2,fe56_3,fe57_1,fe58_1,fe58_2,fe60_1}. 
 Macklin \etal \cite{fe56_3} measured the capture cross sections of $^{56,57}$Fe from 11 to 60 keV. Later on, Allen \etal \cite{fe56_1,fe57_1} used the TOF technique to measure the same from 1 to 800 keV.

Recently, Wang \etal \cite{fe56_2fe57_2} measured the energy averaged $(n,\gamma)$ capture cross sections on $^{56,57}$Fe  from 15 to 90 keV and 11 to 90 keV, respectively, with an error less than 5\%. 
The energy averaged cross sections  on $^{58}$Fe by Allen and Macklin \cite{fe58_2} are extremely scattered and uncertain over the entire range of thermal energies.

 The s-process on iron elements starts from the most abundant $^{56}$Fe and the path uninterruptedly propagates up to $^{59}$Fe. The production of $^{60}$Fe in s-process is governed by the branching at  unstable $^{59}$Fe with $\beta$-decay half-life of 44.495 days.

The short-lived radioisotope (SLRI)  $^{60}$Fe plays the role of an important chronometer for the early solar system (ESS) \cite{fe60_chronometer}. The enrichment in $^{60}$Ni,
 in meteoritic inclusions, is an evidence of its existence in ESS \cite{fe60nienrichment}. Quitt\'{e} \etal \cite{fe60quitte} commented that the
 nucleosynthetic processes (e-process or r-process in neutron-rich environment) that generate $^{62}$Ni should not also produce $^{60}$Ni and hence, it can be a  result from the decay of $^{60}$Fe. 

 The radioactive decays from  $^{60}$Co, the daughter of $^{60}$Fe, are the proof of its existence in the interstellar medium and also are the clear evidence of ongoing neutron capture nucleosynthesis on the pre-existing stable iron isotopes in the massive stars of milky way Galaxy \cite{fe60ongoingspro}.
 Thus the $\gamma$-radioactivity of $^{60}$Fe, similar to the previously discovered radioactive decay of  $^{26}$Al (half-life=7.2$\times$10$^{5}$ years),
can constrain the properties of interstellar medium \cite{fe60_gammaradioactivity}.   Apart from  the study of $\gamma$-radioactivity, $^{60}$Fe$(n,\gamma)$ cross sections are crucial for the study of its formation procedure in various astrophysical sites, such as supernovae and neutron burst nucleosynthesis.

 Uberseder \etal \cite{fe60_1} did the first experiment on the neutron cross section of $^{60}$Fe. They used 47 repeated irradiations and found an average experimental value after summing them up. They folded the experimental neutron energy distribution with the differential Hauser-Feshbach statistical model $(n,\gamma)$ cross section of Ref.~\cite{RTH2000} to obtain a normalization factor. This normalized energy differential cross section is then folded with Maxwell-Boltzmann distribution to obtain the final cross section values for energies ranging from 25 to 100 keV.

Natural cobalt is mono-isotopic. The reaction $^{59}$Co$(n,\gamma)^{60}$Co is important for nuclear dosimetry applications. It is also used as one of the three most common reaction cross-section standards for the experimental techniques of activation. The radioisotope $^{60}$Co is  a major neutron activation product of $^{59}$Co. The neutron activation cycle requires  neutron irradiation without chemical separation and hence, mono-isotopic $^{59}$Co serves as an efficient target. The specific radioactivity of the product is the function of cross sections of both target and product nuclides. 
Fig. \ref {cucongxs} shows the total $(n,\gamma)$ cross sections for $^{59}$Co plotted with experimental data,  taken from  the measurements of Spencer and Macklin \cite{co59_1} and Heil \etal \cite{heil2008}.  The experiment by Macklin \etal \cite{co59_1} was carried out in TOF technique for thermal energies ranging from 2.5 keV to 1 MeV.
   
In Figs. \ref{ningxs1} and \ref{ningxs2}, we have shown the cross sections for $^{58,60-64}$Ni. The experimental 
values are from Refs. \cite{ni58_1,ni58_2,ni58_4,ni58_3ni60_3,ni60stieglitz,ni60perey,ni60corvi,ni61ni62,ni62ni63,ni64,heil2008}.
Elements of nickel are used as important constituents  of structural materials.
The isotope $^{58}$Ni has its origin only from p-process. It is one of the most abundant elements with an isotopic abundance of 68\% and acts as a seed element in the weak s-process nucleosynthesis. The experimental $(n,\gamma)$  cross sections of $^{58}$Ni are taken from Refs. \cite{ni58_1,ni58_2,ni58_4,ni58_3ni60_3}. Perey \etal \cite{ni58_4} presented both energy-averaged and stellar-averaged cross section values for $^{58}$Ni. The energy averaged cross sections are scattered and hence we have not plotted them. The uncertainties  in their measurement  are quoted  as 15\%. There remain sizable differences amongst the existing measurements as well as  evaluated results for $^{58}$Ni cross sections. Guber \etal \cite{ni58_3ni60_3} and Rugel \etal \cite{rugelni58} reported a global decrease.
Most recent measurement by Z\v{u}gec \etal \cite{ni58_1} has used n\_TOF facility at CERN to measure the cross sections for this isotope.
 The presence of significant direct capture component and 
direct-semi-direct  capture component, as suggested in Ref. \cite{dcni58}, has been investigated in Ref.~\cite{ni58_3ni60_3}. %guber
However, most of the experiments are unable to separate out these components.

Experimental cross sections are extremely rare for stable $^{61}$Ni. We have plotted the data of Tomyo \etal \cite{ni61ni62} who provided experimental  values only at three mean energies. We have  taken the experimental  data of  $^{62}$Ni from Refs. \cite{ni62ni63,ni61ni62,ABE08}. Alpizar-Vicente \etal \cite{ABE08}  derived the MACS values after normalizing their measured cross sections with those of Sims and Jhunke \cite{ni62norm_alp_vicente}. Tomyo \etal \cite{ni61ni62} presented  average cross sections for this isotope for energies from 5.5 to 90 keV. They further derived the MACS values by normalizing their data with JENDL-3.3 evaluations \cite{jendl-3.3}. The evaluated cross sections of JENDL-3.3 are multiplied by factors of 2 and 1.5, below 5.5 keV and above  90 keV to derive MACS values. However,  our data is found to underpredict all the measurements for $^{62,63}$Ni.
 
The radioactive isotope $^{63}$Ni ($t_{1/2}$=101 years) is an important branch point nucleus, since, at this point, the reaction flow can be diverted towards $^{63}$Cu/$^{64}$Zn through $\beta$-decay or towards $^{65}$Cu through neutron capture. It is also a long-lived fission product and is used in nuclear transmutation technology. Accurate experimental data are very rare as no natural resource of it is available. We have taken the data from the measurement of Lederer \etal \cite{ni62ni63}. They used n\_TOF facility and determined unresolved cross sections from  10 to 270 keV. The data suffers from a systematic uncertainty of 17\%.  Our results underproduce the cross-section values by an average factor of $\sim$ 2 within the given range of thermal energies.
We have plotted the data from Refs.~\cite{ni64,heil2008} for $^{64}$Ni. Very earlier to the measurement of Heil \etal \cite{heil2008}, H. A. Grench  \cite{ni64} obtained the neutron capture cross sections relative to gold using activation technique and compared their results with Hauser-Feshbach statistical model calculations.

The neutron capture cross sections for $^{63,65}$Cu are shown with experimental data in Fig. \ref{cucongxs}. The  experimental values are from Refs.~\cite{heil2008,cu63_1,cu63_2cu65_1,cu63_3,cu65_2,cu65_3,cu65_4}. 
Recent studies have revealed that major fraction  of the solar copper abundance is produced in massive stars during the s-process. However, a contribution is also believed to come from type-Ia supernova. 
 The element zinc has five stable isotopes.
 They suffer from large propagation effects in abundance distribution, mainly because
 of cross-section uncertainties in $^{63,65}$Cu  and  $^{66,67,68}$Zn \cite{pignatari,heil2008}. 
The isotopes of zinc are also important for galactic chemical evolution study.
The s-process contribution to isotopes of zinc is lower compared to other trans-iron elements (see Fig. 11 of Ref. \cite{pignatari}). The major fractions of  isotopic abundances of  $^{64,66}$Zn are produced during the $\alpha$-rich freeze-out in $\nu$-winds in massive stars. Bisterzo \etal \cite{sr-process} proposed that weak s-process populates mostly the neutron-rich isotopes of zinc. 
The experimental data for $^{64}$Zn and $^{66}$Zn are taken from Jin Xiang \etal \cite{zn64} and Garg \etal \cite{zn66}, respectively, while for $^{68}$Zn, they are taken from Refs.~\cite{zn68_1,zn68_2}. All these are plotted with our calculated results in Fig. \ref{zngangxs}.

\setlength{\tabcolsep}{0.8pt}
\renewcommand{\arraystretch}{1.5}
\begin{table}[htb]
\center
\caption{Maxwellian averaged cross sections at $kT=30$ keV for nuclei near the 
$Z=28$ shell closure. Experimental values are from Ref.~\cite{baorecom,kadonis}. We have also listed theoretical MOST2005 predictions \cite{MOST}.
 For unstable and radioactive nuclei, experimental data are not available. 
 \label{macs30kev}}
\begin{tabular}{crrrcrrr}\hline
 &\multicolumn{3}{c}{MACS (mb)}&
 &\multicolumn{3}{c}{MACS (mb)}\\\cline{2-4}\cline{6-8}
Nucleus&Pres. & Expt. & MOST & Nucleus&Pres. & Expt. & MOST\\
\hline
$_{26}^{56}$Fe&19.0 &11.7$\pm$0.5&36.0&
$_{26}^{57}$Fe&32.1&40$\pm$4&49.6\\
$_{26}^{58}$Fe&10.9&13.5$\pm$0.7&25.1&
$_{26}^{59}$Fe&20.6&&\\
$_{26}^{60}$Fe&3.65&5.15$\pm$1.41&6.8\\
$_{27}^{59}$Co&33.3&39.6$\pm$2.7&53.7&
$_{27}^{60}$Co&46.2&\\
$_{28}^{58}$Ni&42.9&38.7$\pm$1.5&72.2&
$_{28}^{60}$Ni&23.2&29.9$\pm$0.7&39.3\\
$_{28}^{61}$Ni&77.2&82$\pm$8&79.5&
$_{28}^{62}$Ni&11.2&22.3$\pm$1.6&21.2\\
$_{28}^{63}$Ni&32.6& &42.1&
$_{28}^{64}$Ni&5.95&8.0$\pm$0.7&10.0\\
$_{29}^{63}$Cu&76.1&55.6$\pm$2.2&146&
$_{29}^{64}$Cu&128&&\\
$_{29}^{65}$Cu&37.2&29.8$\pm$1.3&48.8\\
$_{30}^{64}$Zn&68.8&59$\pm$5&90.9  &
$_{30}^{65}$Zn&250&&260\\
$_{30}^{66}$Zn&38.1&35$\pm$3&51.0&
$_{30}^{67}$Zn&153 &153$\pm$15&174\\
$_{30}^{68}$Zn&17.9 &19.2$\pm$2.4&20.9&
$_{30}^{70}$Zn&6.03&& 10.1 \\
$_{31}^{69}$Ga&151&139$\pm$6&122&
$_{31}^{71}$Ga&121& 123$\pm$8 &117\\

\hline
\end{tabular}
\end{table}

\renewcommand{\arraystretch}{1.2}
\setlength{\tabcolsep}{6.5pt}
\begin{table*}
\caption{Theoretical MACS values (mb) over a range of energy for reactions with unstable targets $^{59}$Fe, $^{60}$Co, and $^{63}$Ni.
 Experimental values are not available for these nuclei. 
\label{macs_rates}}
\begin{tabular}{ccccccccccccc}
\hline\hline
$kT$ (MeV)&0.005&0.010&0.015&0.020&0.025&0.030&0.040&0.050&0.060&0.080&0.100\\\hline
$^{59}$Fe&81.5&47.0&34.3&27.7&23.5&20.6&16.8&14.4&12.7&10.5&9.1\\
$^{60}$Co&178&105&77.7&62.6&53.0&46.2&37.2&31.4&27.2&21.7&18.2\\
$^{63}$Ni&127&74&55&44&38&33&26&21&18&13&11\\\hline\hline
\end{tabular}
\end{table*}

 Spectroscopic observations reveal that most of the gallium abundances are from s-process
 in massive stars. However, more observational studies are required in order to determine the nucleosynthetic origin of gallium.  
 Gallium has a very low melting point and  high boiling point. Hence, it has the longest liquid range of any metal. Apart from the astrophysical point of view, it is a promising candidate in reactor technology for liquid metallic coolant \cite{ga_coolant}.  
 Nowadays reliable cross sections are also of great demand to study the interaction of gallium with neutrons \cite{ga_neutron_inter}. 
The thermal neutron capture cross sections are taken from Refs.~\cite{ga69_1ga71_2,ga69_2ga71_3,ga69_3ga71_1,ga71_5}. 
The experimental data are extremely old. They were measured more than 30 years ago. We have plotted the data with our theoretical
 results in Fig. \ref{zngangxs}.

\subsection{Maxwellian-averaged cross section (MACS)}
In Table \ref{macs30kev}, we have presented the MACS values for the nuclei shown in the reaction path (Fig. \ref {path}). They are listed with the available experimental values taken 
from the KADoNiS database \cite{kadonis} which is an updated version of recommended values by Bao \etal \cite{baorecom}. 
For the sake of comparison, we have also listed the theoretical MOST2005 calculations \cite{MOST}, whenever available.
   It can be seen that  our theory  reproduces the experimental values, better than MOST2005 calculations, except for a few cases. 

There are  discrepancies in the MACS values of $^{62}$Ni. 
 The direct neutron capture cross section in DWBA calculation by Rauscher and Guber confirmed that there are contributions from   sub-threshold resonance and p-wave capture.
However, thereafter,  Tomyo \etal \cite{ni61ni62} disagreed any p-wave contribution in their measurement. They presented a much larger MACS  of 37$\pm$3.2 mb at 30 keV and claimed that this new large value may solve the longstanding overproduction problem of $^{62}$Ni abundance. %

In Table \ref{macs_rates}, we have presented the Maxwellian-averaged cross sections from 5 to 100 keV for the nuclei, $^{59}$Fe, $^{60}$Co, and $^{63}$Ni. Experimental data are not available for these nuclei. These isotopes  are unstable and hence, are not available for measurement. The isotopes $^{59}$Fe and $^{60}$Co may be subject to weak sr-process \cite{sr-process} in massive stars where they can act as important branch points. Hence, their MACS values would be needed in a complete network calculation to determine the abundances in such astrophysical sites with high neutron density and temperature. The isotope $^{63}$Ni is an important and strong branch-point nucleus, as discussed above. 

\section{Sensitivity to the neutron optical potential}
According to Hauser-Feshbach theory of compound nuclear reactions, total cross section $\sigma_{tot}=\frac{T_{n}T_{\gamma}}{T_tot}$. The transmission coefficients (T) are linearly proportional to average channel widths and hence, $\sigma_{tot} \propto \frac{\Gamma_{n}\Gamma_{\gamma}}{\Gamma_{tot}}$. 
For radiative capture reactions at low energy and for intermediate mass nuclei, average neutron width is much greater than the average radiation width and hence, the resonances in radiative neutron capture is always accompanied with potential elastic scattering. In such a case, $\sigma_{tot}$ would, in principle, be predominantly proportional to $\Gamma_{\gamma}$. Hence, to verify whether the Hauser-Feshbach cross sections depends on the choice of neutron optical potential, we have performed the  calculations with a different  neutron optical  potential based on Jeukene-Lejuene-Mahaux (JLM) interaction \cite{jlm_param}. The other input parameters such as level dnsity, E1 $\gamma$-ray strength function, etc., have been taken from the same references as in case calculations with M3Y potential.

The JLM potential for a given nuclear matter density $\rho_{m}=\rho_{n}+\rho_{p}$ and  asymmetry $\alpha=\frac{\rho_{n}-\rho_{p}}{\rho}$ has been  obtained by folding nuclear matter density distribution with the Reid's hard core nucleon-nucleon interaction.
\begin{equation}
\begin{aligned}
U_{NM}(E)_{\rho \alpha}=\lambda_{v}(E) [V_{0}(E)+\lambda_{V1}\alpha V1(E)]
&&\\+i\lambda_{W}(E) [W_{0}(E)+\lambda_{W1}\alpha W1(E)]
\end{aligned}
\end{equation} 
Where, $\lambda_{V0}$, $\lambda_{V1}$, $\lambda_{W0}$, and $\lambda_{W1}$ are real and imaginary isoscalar and isovector components \cite{jlm_param}.
The final form of the JLM potential considering local density approximation for the application to finite nuclei is given as, 
\begin{equation}
U_{FN}=t(\sqrt\pi)^{-3}\int \frac{U_{NM}(\rho(r\prime),E)}{\rho(r\prime)}exp\frac{{|{\bf r}-\bf{r\prime}|^{2}}}{t_{r}^{2}}\rho(r\prime) d{\bf r\prime}
\end{equation}

\begin{figure}
\includegraphics[scale=0.6]{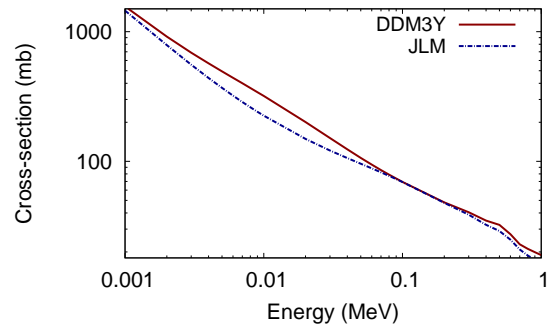}
\caption{Predicted cross section values for $^{69}$Ga$(n,\gamma)^{70}$Ga reaction  with two different density-folded microscopic potentials based on two different nucleon-nucleus interactions. The red solid line represents a calculation with optical model potential based on DDM3Y NN interaction and the blue dotted line represents calculations with JLM optical model potential.
\label{jlm_m3y_ga69}}
\end{figure}
\begin{figure}
\includegraphics[scale=0.5]{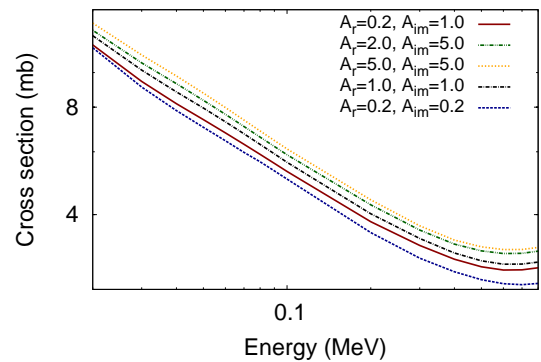}
\caption{Cross sections for the reaction $^{58}$Fe$(n,\gamma)^{59}$Fe with different combinations of real and imaginary potential well depths.
\label{fe58_norm}}
\end{figure}
%We have repeated the calculations with a different microscopic JLM potential while using the same values for other inputs  such as  level densities, strength functions, etc.
 Due to the limitation of the length of the paper, we have shown the comparison between  two potentials only for $^{69}$Ga$(n,\gamma)^{70}$Ga reaction in Fig. \ref{jlm_m3y_ga69}. Two potentials give different results for cross section values. Our microscopic potential has also been found to predict results different from JLM potential in our earlier studies near the $N=82$ and the $N=50$ closed neutron core \cite{n50,n82}. This suggests that the statistical model calculation of  $(n,\gamma)$ cross sections are indeed sensitive to the neutron optical potential.

Further, in order to check the sensitivity to the parameters A$_{r}$ and A$_{im}$ in Eq. \ref{norm_eqn}, we have varied the depths of the  potential for both real and imaginary components by various factors and observed changes in cross section values by different percentages depending upon the reactions concerned. In Fig. \ref{fe58_norm}, we have shown the results for the target $^{58}$Fe for a number of combinations of depth parameters. The depths of real and imaginary parts of the complex potential have been increased and decreased to a fifth to their unnormalized values and the  enhanced or reduced cross section values  are plotted with the cross sections obtained with unnormalized depths. The other reactions more or less follow the same trend.  

It is, therefore, obvious that by properly tuning these parameters for each individual reaction, one can achieve  better agreement with experimental values. Nevertheless, it is evident from the Figs. \ref{fengxs} -\ref{zngangxs} and from presented values in table \ref{macs30kev} that the cross sections can be reasonably described with unnormalized potential depths.  This  was also the case in our previous studies \cite{n82,n50}. Moreover, it is convenient to establish a uniquely parameterized potential model instead of individual fit as it can reflect a more general and global behavior. A further advantage of a single parameter set  is that  the model can subsequently be  applied  to predict the cross section values those are  unknown or yet to be measured.  

\section{Summary}
To summarize, we have performed statistical model Hauser-Feshbach 
calculations in microscopic approach to derive the radiative thermal neutron capture cross sections for nuclei in the vicinity of $Z=28$ proton shell. 
 The nuclei are of astrophysical interests, taking part 
 in the weak component of s-process, occurring in massive stars and also in p- and r-process. The RMF theory is employed to extract target radial densities to use in folding the DDM3Y NN interaction. The $(n,\gamma)$ cross sections are compared with available experimental data and reasonable agreements are achieved for almost  all of the nuclei. This ensures the feasibility of  our theoretical statistical model to predict the radiative thermal capture cross sections, even in the regions where only a few or even no experimental data exist. 
The Maxwellian-averaged cross sections  relevant to  astrophysical applications are presented.

\section{Acknowledgment}
The authors acknowledge the financial assistance from University Grants Commission  (JRF and DRS), Department of Science and Technology, and Alexander Von Humboldt Foundation.


\begin{thebibliography}{99}
\bibitem{weak_sp}J. G. Peters, Astrophys. J. {\bf154}, 225 (1968).
\bibitem{psr}K. Farouqi, K. -L. Kratz, and B. Pfeiffer, Publications of the Astronomical Society of Australia {\bf26}, 194 (2009).
\bibitem{sr-process} S. Bisterzo, R. Gallino, M. Pignatari, L. Pompeia, K. Cunha, and V. Smith, Memorie della Societ\'{a} Astronomica Italiana
 {\bf75}, 741 (2004).
\bibitem{fe60_1}E. Uberseder, R. Reifarth, D. Schumann, I. Dillmann, C. D. Pardo, J. Gorres, M. Heil, F. Kappeler, J.Marganiec, J.  Neuhausen, M. Pignatari, F. Voss, S. Walter, and M. Wiescher, Physical Review Letters {\bf102}, 151101 (2009).
\bibitem{book}Chemical Abundances and Mixing in Stars in the Milky Way and its Satellites: Proceedings of the ESO-Arcetric Workshop held in Castiglione della Pescaia, Italy, 13-17 Sept. 2004, S. Randich and L. Pasquini, Springer-Verlag, Berlin, Heidelberg, 2006, printed in Netherlands.
\bibitem{galactic} A. Alib\`{e}s, J. Labay, and R. Canal,  Astronomy \& Astrophysics, {\bf370}, 1103 (2001).
\bibitem{talys}A. J. Koning, S. Hilaire, and M. Duizvestijn, in {\em Proceedings
of the International Conference on Nuclear Data for Science
and Technology, April 22-27, 2007, Nice, France}, edited by
O. Bersillon, F. Gunsing, E. Bauge, R. Jacqmin, and S. Leray
(EDP Sciences, Cedex, France, 2008), p. 211.\textless www.talys.eu\textgreater.
\bibitem{density_factor} D. N. Basu, J. Phys. G {\bf30}, B7 (2004).
\bibitem{n82}S. Dutta, D. Chakraborty, G. Gangopadhyay, and A. Bhattacharyya, Physical Review C {\bf93}, 024602 (2016).
\bibitem{n50}S. Dutta, G. Gangopadhyay, and Abhijit Bhattacharyya, Physical Review C {\bf 94}, 024604 (2016). 
\bibitem{iliadis}See, for example, C. Iliadis, Nuclear Physics of Stars (Wiley-
VCH Verlag GmbH \& Co., Boschstr., Weinheim, Germany, 2007).
\bibitem{ddm3y}A. M. Kobos, B. A. Brown, P. E. Hodgson, G. R. Satchler, and A. Budzanowski, Nucl. Phys. A {\bf384}, 65 (1982).
\bibitem{fsugold}B. G. Todd-Rutel and J. Piekarewicz, Phys. Rev. Lett. {\bf95}, 122501 (2005).
\bibitem{55-60}S. Dutta, D. Chakraborty, G. Gangopadhyay, and A. Bhattacharyya, Phys. Rev. C {\bf91}, 025804 (2015).
\bibitem{60-80}C. Lahiri and G. Gangopadhyay, Eur. Phys. J. A {\bf47}, 87 (2011).
\bibitem{80-90}C. Lahiri and G. Gangopadhyay, Phys. Rev. C {\bf84}, 057601 (2011).
\bibitem{110-125}D. Chakraborty, S. Dutta, G. Gangopadhyay, and A. Bhattacharyya, Phys. Rev C {\bf91}, 057602 (2015).
\bibitem{40-55}D. Chakraborty,S. Dutta, G. Gangopadhyay, and Abhijit Bhattacharyya, Phys. Rev. C {\bf94}, 015802 (2016).
\bibitem{strength_4} S. Goriely, E. Khan, M. Samyn, Nucl. Phys. A {\bf739}, 331 (2004).
\bibitem{ldmodel_4}S. Goriely, S. Hilaire and A. J. Koning, Phys. Rev. C {\bf78}, 064307 (2008).
\bibitem{boson_part_func}S. Hilaire, J. P. Delaroche, and M. Girod, Eur. Phys. J. A {\bf12}, 169 (2001).
\bibitem{hfb_gr}S. Goriely, M. Samyn, and J. M. Pearson, Phys. Rev. C {\bf75}, 064312 (2007).
\bibitem{bsk14} S. Hilaire, S. Goriely, Nucl. Phys. A {\bf779} (2006) 63; S. Goriely, S. Hilaire, A. J. Koning, Phys. Rev. C {\bf78} (2008) 064307.
\bibitem{devries}H. DeVries, C. W. DeJager, and C. DeVries, Atomic Data Nuclear Data Tables {\bf36}, 495 (1987).
\bibitem{angeli}I. Angeli, At. Data Nucl. Data Tables {\bf 87}, 185 (2004).
\bibitem{heil2008}M. Heil, F. K{\"a}ppeler, E. Uberseder, R. Gallino, M. Pignatari, Phys. Rev. C {\bf77}, 015808 (2008).%co59plot
\bibitem{fe56_1}B. J. Allen, A. R. de L. Musgrove, J. W. Boldeman, M. J. Kenny, and R. L. Macklin, Nucl. Phys. A  {\bf269}, 408 (1976).
\bibitem{fe56_2fe57_2}T. Wang, M. Lee, G. Kim, Y. Oh, W. Namkung, T.-I. Ro, Y. -R. Kang, M. Igashira, and T. Katabuchi, Nucl. Instrum. Methods in Physics Res. B {\bf268}, 440 (2010).
\bibitem{fe56_3}R. L. Macklin, P. J. Pasma, and J. H. Gibbons, Physical Review {\bf136}, B695 (1964).
\bibitem{fe57_1}B. J. Allen, A.R. de L. Musgrove, R. Taylor, and R. L. Macklin, Meet. on Neutr. Data of Struct. Mat., Geel 1977, p.476 (1977).
\bibitem{fe58_1}Yu. N. Trofimov, Atomnaya Energiya {\bf58}, 278 (1985).
\bibitem{fe58_2}B. J. Allen and R. L. Macklin,  Jour. of Physics G: Nucl.and Part.Phys. {\bf6}, 381 (1980).
\bibitem{fe60_chronometer}A. Boss, Astrophys. J. {\bf660}, 1707 (2007).
\bibitem{fe60nienrichment}J. Birck and G. Lugmair, Earth Planet. Sci. Lett. {\bf90}, 131
(1988).
\bibitem{fe60quitte}G. Quitt\'{e}, A. N. Halliday, B. S. Meyer, and A. Markowski, Astrophys. J. {\bf655}, 678 (2007).
\bibitem{fe60limongi}M. Limongi and A. Chieffi, The Astrophysical Journal {\bf647}, 483 (2006).
\bibitem{fe60ongoingspro}G. Rugel, T. Faestermann, K. Knie, G. Korschinek,  M. Poutivtsev, D. Schumann, N. Kivel, I. Gunther-Leopold, R. Weinreich, and M. Wohlmuther, Phys. Rev. Lett. {\bf 103}, 072502 (2009).
\bibitem{fe60_gammaradioactivity}R. Diehl, N. Prantzos, and P. von Ballmoos, Nucl. Phys. A {\bf777}, 70 (2006).
\bibitem{RTH2000} T. Rauscher and F.-K. Thielemann, Atomic Data Nuclear Data Tables {\bf 75}, 1 (2000). 
\bibitem{co59_1}R. R. Spencer and R. L. Macklin, Nuclear Science and Engineering {\bf61}, 346 (1976).
\bibitem{ni58_1}P. Z\v{u}gec, Physical Review C  {\bf89}, 014605 (2014).
\bibitem{ni58_2}K. Wisshak, F. K{\"a}ppeler, G. Reffo, and F. Fabbri, Nuclear Science and Engineering {\bf86}, 168 (1984).
\bibitem{ni58_4}C. M. Perey, J. A. Harvey, R. L. Macklin, R. R. Winters, and F. G. Perey, Oak Ridge National Lab. Reports No.5893 (1982).
\bibitem{ni58_3ni60_3}K. H. Guber, H. Derrien, L. C. Leal, G. Arbanas, D. Wiarda, P. E. Koehler, and J. A. Harvey, Physical Review C {\bf82}, 057601 (2010).
\bibitem{ni60stieglitz} R. G. Stieglitz, R. W. Hockenbury, R. C. Block, Nuclear Physics A {\bf163}, p.592 (1971)
\bibitem{ni60perey}C. M. Perey, J. A. Harvey, R. L. Macklin, R. R. Winters, and F. G. Perey,  Oak Ridge National Lab. Reports, No.5893 (1982).
\bibitem{ni60corvi}F. Corvi, G. Fioni, F. Gunsing, P. Mutti, and L. Zanini,  Nuclear Physics A {\bf697}, 581 (2002).
\bibitem{ni61ni62}A. Tomyo, Y. Temma, M. Segawa, Y. Nagai, H. Makii, T. Shima, T. Ohsaki, and M. Igashira,
 Astrophysical Journal  {\bf623}, L153 (2005).
\bibitem{ni62ni63}C. Lederer {\em et al.}, Physical Review C {\bf89}, 025810 (2014).
\bibitem{ni64}H. A. Grench, Physical Review {\bf140}, B1277 (1965).
\bibitem{rugelni58}G. Rugel, I. Dillmann, T. Faestermann {\em et al.}, Nucl. Instrum.
Methods B {\bf259}, 683 (2007).
\bibitem{dcni58}W. E. Parker, M. B. Chadwick, F. S. Dietrich {\em et al.}, Phys. Rev.
C {\bf52}, 252 (1995).
\bibitem{ABE08}A. M. Alpizar-Vicente, T. A. Bredeweg, E. -I. Esch, U. Greife, R. C. Haight, R. Hatarik, J. M. O'Donnell, R. Reifarth, R. S. Rundberg, J. L. Ullmann, D. J. Vieira, and J. M. Wouters, Phys. Rev. C {\bf77}, 015806 (2008).
\bibitem{ni62norm_alp_vicente}G. H. E. Sims and D. G. Juhnke, J. Inorg. Nucl. Chem. {\bf 32}, 411 (1970).
\bibitem{jendl-3.3}K. Shibata {\em et al.}, J. Nucl. Sci. Technol. {\bf39} (2002) 1125. \textless http://wwwndc.jaea.go.jp/jendl/j33/j33.html\textgreater.
\bibitem{cu63_1}V. A. Tolstikov, V. P. Koroleva, V. E. Kolesov, and A. G. Dovbenko, Atomnaya Energiya {\bf21}, 45 (1966).
\bibitem{cu63_2cu65_1} G. G. Zaikin, I. A. Korzh, N. T. Sklyar, and I. A. Totskii, Soviet Atomic Energy {\bf25}, 1362 (1968).
\bibitem{cu63_3}J. M. Blair, M. Deutsch, K. M. Griesen, A. O. Hanson, G. A. Linenberger, J. A. Miskel, R. F. Taschek, C. M. Turner, and J. H. Williams, Los Alamos Scientific Lab. Reports, No.95 (1944).
\bibitem{cu65_2}V. A. Tolstikov, V. E. Kolesov, A. G. Dovbenko, and Ju. Ja. Stavisskij, Atomnaya Energiya  {\bf17}, 505 (1964).
\bibitem{cu65_3}Yu. Ya. Stavisskiy and V. A. Tolstikov, Atomnaya Energiya {\bf10}, 508 (1961).
\bibitem{cu65_4}A. E. Johnsrud, M. G. Silbert, and H. H. Barschall, Physical Review {\bf116}, 927 (1959).
\bibitem{pignatari}M. Pignatari, R. Gallino, M. Heil, M. Weischer, F. K\"{a}ppeler, F. Herwig, and S.Bisterzo, The Astrophys. Jour. {\bf710}, 1557 (2010).
\bibitem{zn64}Chen Jinxiang, Shi Zhaomin, Tang Guoyou, and Zhang Guohui, Chinese Journal of Nuclear Physics (Beijing)., {\bf17}, 342 (1995).
\bibitem{zn66}J. B. Garg, V. K. Tikku, J. A. Harvey, R. L. Macklin, J. Halperin, Physical Review C {\bf24}, 1922 (1981).
\bibitem{zn68_1}J. B. Garg, V. K. Tikku, J. A.  Harvey, J. Halperin, and R. L. Macklin, Physical Review C {\bf25}, 1808 (1982).
\bibitem{zn68_2}A. I. Leipunskiy, O. D. Kazachkovskiy, G. Ja. Artyukhov, A. I. Baryshnikov, T. S. Belanova, V. I. Galkov, Yu. Ja. Stavisskiy, E. A. Stumbur, and L.E.Sherman, Second Internat. At. En. Conf., Geneva  {\bf15}, 50 (1958).
\bibitem{baorecom}Z. Y. Bao,  H. Beer, F. K{\"a}appeler, F. Voss, and K. Wisshak, Atomic Data and Nuclear Data Tables {\bf 76}, 70 (2000).
\bibitem{kadonis} I. Dillmann, R. Plag, F. K\"{a}ppeler, T. Rauscher,
KADoNiS v0.3 - The third update of the `Karlsruhe
Astrophysical Database of Nucleosynthesis in Stars'
in {\em EFNUDAT Fast Neutrons, Proceedings of the  Scientific Workshop on Neutron 
Measurements, Theory, and Applications}
28 - 30 April 2009,  Geel, 
Belgium, edited by F. -J. Hambsch (Publications Office of the European Union, Luxembourg, 2010) p. 55.
\textless www.kadonis.org.\textgreater.
\bibitem{MOST}M. Arnould and S. Goriely, Phys. Rep. {\bf 384}, 1 (2003).
\bibitem{ga_coolant}T. Sawada, A. Netchaev, H. Ninokata, H. Endo, Progress in Nuclear Energy {\bf37}, 313 (2000).
\bibitem{ga_neutron_inter}L. Koester, K. Knopf, W. Waschkowski, and A. Kl{\"u}ver, Zeitschrift f{\"u}r Physik A Atoms and Nuclei {\bf318}, 347 (1984).
\bibitem{ga69_1ga71_2}A. G. Dovbenko, V. E.  Kolesov, V. P. Koroleva, and V. A. Tolstikov, Soviet Atomic Energy {\bf26}, 82 (1969).
\bibitem{ga69_2ga71_3}G. G. Zaikin, I. A. Korzh, M. V. Pasechnic, and N. T. Skljar,  Ukrainskii Fizichnii Zhurnal {\bf16}, 1205 (1971).
\bibitem{ga69_3ga71_1}Yu. Ya. Stavisskii and V. A. Tolstikov, Atomnaya Energiya {\bf7}, 259 (1959).
\bibitem{ga71_5}A. E. Johnsrud, M. G. Silbert, and H. H. Barschall, Physical Review  {\bf116}, 927 (1959).
\bibitem{jlm_param}J.-P. Jeukenne, A. Lejeune, and C. Mahaux, Phys. Rev. C {\bf16}, 80 (1977).




\end{thebibliography}
\end{document}